\documentclass[pdflatex,sn-mathphys-num]{sn-jnl}%

\usepackage{graphicx}%
\usepackage{multirow}%
\usepackage{amsmath,amssymb,amsfonts}%
\usepackage{amsthm}%
\usepackage{mathrsfs}%
\usepackage[title]{appendix}%
\usepackage{xcolor}%
\usepackage{textcomp}%
\usepackage{manyfoot}%
\usepackage{booktabs}%
\usepackage{algorithm}%
\usepackage{algorithmicx}%
\usepackage{algpseudocode}%
\usepackage{listings}%
\usepackage{svg}%
\graphicspath{{figures/}}

\theoremstyle{thmstyleone}%
\newtheorem{theorem}{Theorem}%
\newtheorem{corollary}[theorem]{Corollary}%
\newtheorem{lemma}[theorem]{Lemma}%

\theoremstyle{thmstyletwo}%

\theoremstyle{thmstylethree}%

\begin{document}

\title[Influencer's Games in Resource Landscapes]{Learning Dynamics of the Influencer's Game in Resource Landscapes}

\author*[1]{\fnm{Mark} \sur{Lovett}}\email{mark.lovett.gr@dartmouth.edu}

\author[1]{\fnm{Feng} \sur{Fu}}\email{feng.fu@dartmouth.edu}

\affil*[1]{\orgdiv{Mathematics}, \orgname{Dartmouth College}, \orgaddress{\street{29 North Main Street}, \city{Hanover}, \postcode{03755}, \state{NH}, \country{USA}}}

\abstract{ Across domains as varied as ecology, economics, and online platforms, individuals are incentivized to position themselves within resource-rich strategy spaces. In this work we introduce a new spatial game framework called the influencer's game. Within this new framework players' strategies are represented as positions within a trait space. Players then receive a share of a portion of a resource based on their relative influence at that resource's points traits. For the influencer's game, we show that there is a unique symmetric Nash equilibrium that exists under strict log-concavity of the influence kernel. Using the unique equilibrium we show that the stability of the system depends on the influence kernel's parameters, number of players, and the resource distribution shape. For any dimension we demonstrate a stability threshold that depends on both the scatter and curvature of the players. When this threshold is crossed, the symmetric equilibrium destabilizes and gives way to a variety of bifurcation types; including, specialist clusters, limit cycles, and sub-games that reproduce the larger game's structure. We ground these results using real-world data for resources as a demonstration of the game's application. Finally we extend the game dynamics to use multi-agent reinforcement learning and further show that the predicted bifurcation structure persists under partial observability and discretized strategy spaces. In all, the influencer's game provides a novel framework for predicting when competitive pressure drives convergence versus fragmentation with possible applications across domains.}


\maketitle

\section{Introduction}\label{sec1}

When individuals compete for resources (whether firms courting consumers, species foraging across a landscape, or political candidates chasing voter support), they must interact with others and try to predict the best traits to achieve the most competitive advantage. The dilemma that players face is to crowd around resource-rich regions with increased competition, or to explore niches that risk scarcity but may hold abundant, uncontested resources. The tension between crowding and opportunity shapes niche formation in ecology \cite{vandermeer1972niche,tilman1982resource}, product differentiation in economics \cite{hotelling1929stability}, and platform positioning in politics \cite{downs1957economic,enelow1984spatial}. There is a great deal of prior work on spatial competition. The groundbreaking economic model of spatial competition was Hotelling's model of duopoly \cite{hotelling1929stability,eaton1975principle,salop1979monopolistic,irmen1998competition} which has two sellers choose locations on a line of consumers. The sellers compete for the consumers based on the distance to the consumer, so that consumers always choose sellers that are closest to them. This results in the sellers moving to the center of the consumer distribution, in the case of the line the mean. This was coined the principle of minimum differentiation. Later the principle of minimum differentiation was found to be very sensitive to settings such as varied consumer preference types and player counts, which gave way to new spatial competition games like facility-location games \cite{owen1998strategic,drezner2004facility}. Additional spatial competition games include Blotto games \cite{borel1953theory,roberson2006colonel,hart2008discrete,roberson2012non} and Tullock contests \cite{tullock1980efficient,skaperdas1996contest,ghosh2023best}. Blotto games center around dividing a fixed budget across discrete battlefields and Tullock contests model competitive effort with a prize awarded with probability proportional to the effort committed. In the political science context, spatial competition has been used to study the dynamics of voters and candidates, which has given rise to both the median and mean voter theory \cite{caplin1991aggregation,schofield2007mean,jones2022polarization}. While in ecology, spatial competition has been represented as the competition for shared resources such as food, water, and territory \cite{chesson2000mechanisms,fort2020combining,levin2009princeton}. Each of the above preexisting frameworks was developed for a particular setting or limited in how broad its results could be generalized.

Here we introduce the influencer's game, a framework that generalizes and connects preexisting approaches by allowing $N$ players to choose strategic positions in an $L$-dimensional strategy space $\mathbb{D}$. Each player exerts an influence kernel, a function that determines how strongly they compete for a resource at any given resource trait $b$, and receives a share of a resource distribution $B(b)$ over a resource trait space $\mathbb{B}$ via a proportional allocation rule. We illustrate the framework on a resource landscape derived from LinkedIn job postings from 2023--2024. Below a critical depth (introduced later), the symmetric equilibrium destabilizes and a cascade of bifurcations follows. The result is that different specialist niches form at different depths.

What makes a symmetric influencer's game unique compared to many existing frameworks is that it remains analytically tractable at any number of players. This is primarily because the Jacobian at the symmetric equilibrium decomposes into two $L\times L$ eigenproblems whose size does not grow with $N$ (Theorems~\ref{thm:block_jacobian} and~\ref{thm:spectral}; Appendix~\ref{app:theorems}). The stability of the symmetric equilibrium is then decided by a single condition that compares how the resource pulls on a player with how sharply the player's influence kernel's curvature around its position (Theorem~\ref{thm:general_stability}). For two players this quantity vanishes, so with strictly log-concave influence kernels the symmetric equilibrium is locally stable under gradient-ascent dynamics regardless of dimensionality (Theorem~\ref{thm:2playerstability}); in this regime, two competitors face a consistent incentive to cluster at the symmetric equilibrium.

In the Gaussian influence kernel case, we find that when each player's competitive reach ($\sigma$) is broad, all players converge to the resource-weighted mean, recovering the classical principle of minimum differentiation \cite{caplin1991aggregation,schofield2007mean}. As the competitive reach $\sigma$ decreases below a critical threshold $\sigma_0^*$, the symmetric equilibrium destabilizes and a cascade of bifurcations produces specialist clusters around the peaks of the resource distribution. In the extreme limit $\sigma \to 0$, two-player Gaussian influencer's games with a unimodal, symmetric resource distribution converge to winner-takes-all allocation at the median of the distribution, recovering the classical median voter theorem \cite{downs1957economic,enelow1984spatial} as a special case (SI Section~5.2.2). The same bifurcation structure persists under more realistic learning conditions: independent Q-learning agents with partial observability and discretized strategy spaces reproduce the same sequence of equilibrium types, and the theory extends naturally to Beta, multivariate Gaussian, and Dirichlet kernels on multi-dimensional and simplex strategy spaces.

\begin{figure*}[t]
\centering
\includegraphics[width=\textwidth]{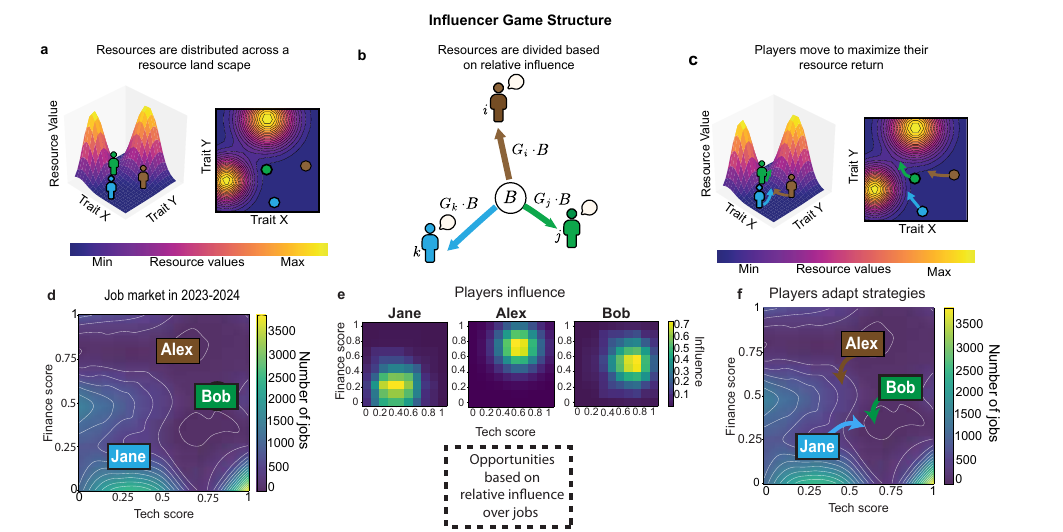}
\caption{\textbf{The influencer's game: a unifying framework for niche competition.}
    \textbf{a},~Three agents ($i$, $j$, $k$) choose strategic positions in a resource trait space $\mathbb{B}$ over which a resource distribution $B(b)$ is defined (e.g.\ consumer demand, voter support, or ecological resources).
    \textbf{b},~Proportional allocation: each agent $s=i,j,k$ receives a share $G_s \cdot B$ of the resource $B(b)$ proportional to its relative influence ($\sum_s G_s = 1$), yielding a constant-sum payoff across all players $\sum_{s\in N} u_s(\mathbf{x})=u$.
    \textbf{c},~Agents adapt strategic positions via a learning mechanism.
    \textbf{d},~Motivating example: LinkedIn job postings (2023--2024) scored on $[0,1]$ for technology and finance relevance. Job seekers compete for postings by choosing strategic positions in the resource trait space.
    \textbf{e},~Each job seeker exerts a multivariate Gaussian influence $f_i(x_i, b)$ determining their proportional share of job opportunities via $G_i(\mathbf{x}, b)$; proximity to a resource trait yields greater influence and more opportunities.
    \textbf{f},~Arrows show how job seekers adjust strategic positions to maximize expected opportunities via adaptive dynamics or reinforcement learning.}\label{fig:framework}
\end{figure*}

\section{Results}
\paragraph{Influencer's Game Model}
    In Figure~\ref{fig:framework} we introduce the influencer's game, a model of spatial competition in which $N$ players choose strategic positions in an $L$-dimensional strategy space $\mathbb{D}$ to compete for resources in a resource distribution $B(b)$ which is defined over a resource trait space $\mathbb{B}$. Resource trait spaces like the one shown in Figure~\ref{fig:framework}\textbf{a} are ubiquitous across competitive settings: in social media, the trait $b$ may represent a topic or sentiment score over which consumer demand is distributed; in politics, $b$ may be an ideological score carrying voter support; and in ecology, $b$ may represent beak size or food preference indexing available resources.
 Across settings, players must then decide where to position themselves strategically in order to capture a share of the resource. We represent this share via a proportional allocation rule (Figure~\ref{fig:framework}\textbf{b}): each player is allocated a share of the resource at trait $b$ proportional to their relative influence $G_i(\mathbf{x}, b)$, where $\mathbf{x} = (x_1, \ldots, x_N) \in \mathbb{D}^N$ is the tuple of strategic positions, with each $x_i \in \mathbb{D} \subseteq \mathbb{R}^L$; in one dimension ($L=1$) each $x_i$ is a scalar and $\mathbf{x} \in \mathbb{R}^N$. We define the allocation rule $G_i(\mathbf{x}, b)$ as the player's relative influence $f_i(x_i, b)$ at trait $b$ compared to the total influence exerted by all players at that trait,
    \begin{equation}\label{eq:prop}
        G_i(\mathbf{x}, b) = \frac{f_i(x_i, b)}{\sum_{j=1}^N f_j(x_j, b)}.
    \end{equation}
    What the influence function stands for depends on the application. For a political candidate it is how well the campaign mobilizes the voters who sit at ideological position $b$ \cite{coughlin1992probabilistic,enelow1984spatial}; for a species it is the competitive pressure it puts on the food or mates available under condition $b$ \cite{macarthur1967limiting,vandermeer1972niche}.

    The payoff to each player is then the expected share of the resource they receive across all traits, given by
    \begin{equation}\label{eq:payoff}
        u_i(\mathbf{x}) = \sum_{b \in \mathbb{B}} B(b) G_i(\mathbf{x}, b).
    \end{equation}
    Players adapt their strategic positions over time to maximize payoffs, for example via gradient ascent on the payoff landscape or reinforcement learning (Figure~\ref{fig:framework}\textbf{c},\textbf{f}).

\begin{figure*}[t]
\centering
\includegraphics[width=\textwidth]{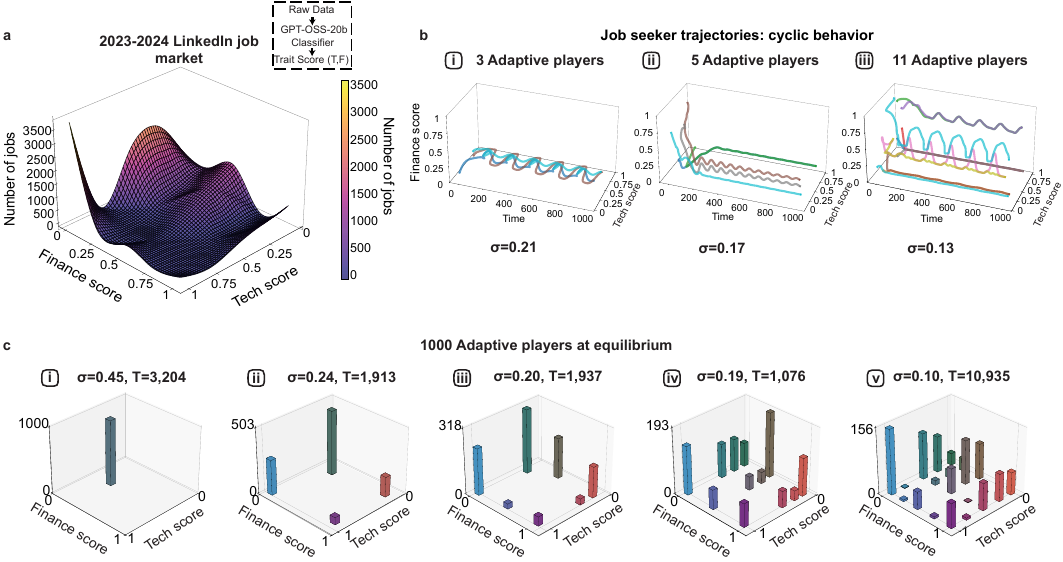}
\caption{\textbf{Rich strategy dynamics of applicants in competitive job markets}
    \textbf{a},~LinkedIn job postings (2023--2024) scored along technology and finance relevance using a language model (LM) GPT-OSS-20B, yielding an empirical resource landscape $B(b)$ over the resource trait space $\mathbb{B} = [0,1]^2$. Peaks represent clusters of similar job opportunities; valleys indicate resource-poor regions.
    \textbf{b},~Cyclic dynamics for $N \in \{3,\,5,\,11\}$ symmetric job seekers (left to right) with multivariate Gaussian influence kernels ($\Sigma=\sigma^2 I_{2}$)
under adaptive dynamics $\nabla_t x_i = \nabla_{x_i} u_i(\mathbf{x})$. As $N$ increases, the competitive reach $\sigma$ must decrease for cycles to appear.
    \textbf{c},~Equilibrium densities for $N = 1{,}000$ players at competitive reach values $\sigma \in \{0.45,\, 0.24,\, 0.20,\, 0.19,\, 0.10\}$ (left to right). As competitive reach narrows ($\sigma\to 0$), players cluster around modes of the job-posting distribution, mirroring specialization dynamics in ecology and social-media niches.}\label{fig:linkedin}
\end{figure*}

    We motivate the influencer's game with an empirical model of job market dynamics. In Figures~\ref{fig:framework} and~\ref{fig:linkedin} we present the spatial resource distribution of job postings from 2023--2024 \cite{arsh_koneru_2024}. Each posting is scored on a 0--1 scale for technology and finance. This results in two separate trait scores for each posting, giving the posting a unique position. Repeating this over all postings and grouping postings based on their trait scores yields a two-dimensional resource landscape of job postings. We can then interpret the resulting influencer's game as players (job seekers), their strategic positions (skill profiles), and the resources (number of job postings). Then each seeker can capture potential postings through their influence $f_i(x_i, b)$ which is the distance between the seeker's skill profile and the posting's trait score. Thus, job seekers are incentivized to develop skill profiles that are close to the posting traits that have dense job opportunities while keeping in mind competition from other seekers.

    We model each seeker's influence with a multivariate Gaussian kernel
    \begin{equation}\label{eq:MV_gaussian_kernel}
        f_i(x_i, b) = \exp\left(-\frac{1}{2}(b - x_i)^\top \Sigma^{-1} (b - x_i)\right)
    \end{equation}
    with a common isotropic covariance matrix $\Sigma = \sigma^2 I_2$ (the two-by-two identity matrix scaled by $\sigma^2$), centered at each seeker's strongest trait $x_i$. In the job-market setting, the isotropic form of the covariance matrix makes a job seeker's technology and finance skills equally transferable. Thus, sharing $\Sigma$ across the population extends this assumption to every job seeker. We therefore interpret $\sigma$ as a job seeker's competitive reach, or how broadly their skills apply across job postings. Covariance matrices that are not isotropic allow reach to vary between the technology and finance directions, and these are addressed in the analytical extensions below.
 In Figure~\ref{fig:linkedin}\textbf{b} we see that under adaptive dynamics (gradient ascent on the payoff landscape such that $\nabla_t x_i = \nabla_{x_i} u_i(\mathbf{x})$), three independent job seekers subject to the same job market demonstrate cyclic behavior. As the number of job seekers increases ($N=3,5,11$), similar cyclic behavior persists. In both the 11- and 5-job-seeker cases, the cyclic behavior is more prominent for some players than others, with some players converging to a clustered equilibrium of varying numbers of players.

    Observing the behavioral differences between groups of players in Figure~\ref{fig:linkedin}\textbf{b}, it is natural to ask whether similar clustering or cycles occur at higher numbers of players. In Figure~\ref{fig:linkedin}\textbf{c} there are now $1{,}000$ players who are independently competing for jobs, and we see that as the competitive reach $\sigma$ decreases, players cluster at different levels, with more clusters appearing as their competitive reach decreases further. In Figure~\ref{fig:linkedin}\textbf{c.i}--\textbf{c.v}, as the competitive reach $\sigma$ decreases through $\sigma \in \{0.45,\, 0.24,\, 0.20,\, 0.19,\, 0.10\}$, corresponding to increasingly specialized skill sets, players cluster around the peaks of the job-posting distribution, mirroring niche partitioning in ecology \cite{fort2020combining,dieckmann1999origin} and echo-chamber formation on social media \cite{cinelli2021echo}.
As $\sigma$ decreases, larger groups of players bifurcate and more smaller groups emerge, producing a hierarchical organization of clusters: clusters formed at one value of $\sigma$ themselves split into smaller clusters as $\sigma$ decreases further, each behaving like a smaller influencer's game (we use ``hierarchical'' in this sense throughout). Moreover, the clusters that emerge are not all the same size, with some clusters having more players than others, and as we approach a smaller $\sigma$ the players begin to approximate the shape of the resource distribution.

The behavior in Figure~\ref{fig:linkedin} raises three theoretical questions: when does a symmetric Nash equilibrium exist; when is it unique; and when is it stable, i.e.\ when do the learning dynamics converge to it? We address each in turn. For symmetric influencer's games with quasi-concave influence kernels on a compact, convex domain $\mathbb{D}$, a standard fixed-point argument \cite{debreu1952social,reny2020nash} guarantees at least one symmetric pure-strategy Nash equilibrium (Theorem~\ref{thm:existence}; Appendix~\ref{app:theorems}; proof in SI Section~2.1). When the kernel is strictly log-concave (its logarithm is strictly concave), a condition satisfied by the Gaussian and multivariate Gaussian kernels, by the Beta kernel for $\sigma \in (0, \tfrac{1}{2})$, and by the Dirichlet kernel for every $\sigma > 0$ (SI Sections~3.3 and~4.2; each is parameterized by a competitive-reach parameter analogous to $\sigma$), the symmetric equilibrium is unique (Theorem~\ref{thm:uniqueness}; proof in SI Section~2.2). Existence tells us there is always a symmetric equilibrium to talk about, whatever $N$, the dimension, or the shape of the resource. Neither uniqueness nor existence of the symmetric equilibrium depends on the competitive reach $\sigma$ or the player count $N$; however, the stability of the symmetric equilibrium does depend on these parameters.

The sign of the real part of the maximal eigenvalue of the Jacobian determines when the symmetric equilibrium will be stable under the gradient-ascent dynamics. We can use the symmetry of the game to reduce the maximal eigenvalue problem into two $L\times L$ blocks, $A^{+}$ for perturbations that move all players together and $A^{-}$ for perturbations that move them apart (Theorems~\ref{thm:block_jacobian} and~\ref{thm:spectral}). Both $A^+,A^-$ depend on only two matrices, the expected curvature of the log-kernel at the equilibrium, $C = -\mathbb{E}_B[\nabla_x^2 \ln f(x^*,b)]$, and the scatter of its gradient, $S = \mathbb{E}_B[\nabla_x \ln f(x^*,b)\,\nabla_x \ln f(x^*,b)^{\top}]$. The curvature $C$ describes the sharpness of a player's influence kernel around their position. The scatter $S$ is the strength of the incentivization for the players to seperate into different directions. Importantly, under symmetry, moving together never destabilizes the players. Since $A^{+} = -\frac{N-1}{N^2}C$ is negative definite whenever $C$ is positive definite, which is satisfied by a log-concave $f$, the equilibrium only loses stability through the players separating. This occurs when
\begin{equation}\label{eq:general_depth}
    \rho = \frac{N-2}{N-1}\,\lambda_{\max}\!\left(C^{-1/2}\,S\,C^{-1/2}\right) > 1,
\end{equation}
whereas for $\rho < 1$ the symmetric Nash equilibrium is locally stable (Theorem~\ref{thm:general_stability}; proof in SI Section~2.3.3). The symmetric equilibrium thus loses stability once the pull of the resources in its strongest direction outweighs the curvature of the kernel.

The simplest case of this is the two-player game. For $N = 2$ the factor $\frac{N-2}{N-1}$ vanishes, so that the depth $\rho = 0$ indicating that it is impossible for 2 symmetric players with log-concave kernels to have an unstable symmetric equilibrium (Theorem~\ref{thm:2playerstability}). Two competitors therefore always return to the shared position under local gradient-ascent learning. Thus for instability and bifurcations to occur there need to be at least 3 players or asymmetric kernels. This is reminiscent of the median voter and minimum differentiation principles that we discussed in the introduction, but expanded to multiple dimensions and kernels. When players' numbers grow the factor approaches one, so the symmetric equilibrium stability becomes more difficult to maintain and depends on the influence parameters and the resource distribution.

When in one dimension $C$ and $S$ are scalars, and the condition $\rho < 1$ takes a simpler form by reducing to the inequality
\begin{equation}\label{eq:1d_stability}
    \sum_{b \in \mathbb{B}} B(b) d'(b)<-\frac{N-2}{N-1}\sum_{b \in \mathbb{B}} B(b) d(b)^2,
\end{equation}
where $d(b)=\partial_{x_i} \ln f_i(x^*, b)$ and $d'(b) = \partial_{x_i}^2 \ln f_i(x^*, b)$. This condition combines the kernel geometry, the resource distribution, and the number of players. This allows us to easily apply this rule to one-dimensional games as we show later.

\paragraph{Dynamics in One-Dimensional Gaussian Influencer's Games}

\begin{figure*}[t]
\centering
\includegraphics[width=\textwidth]{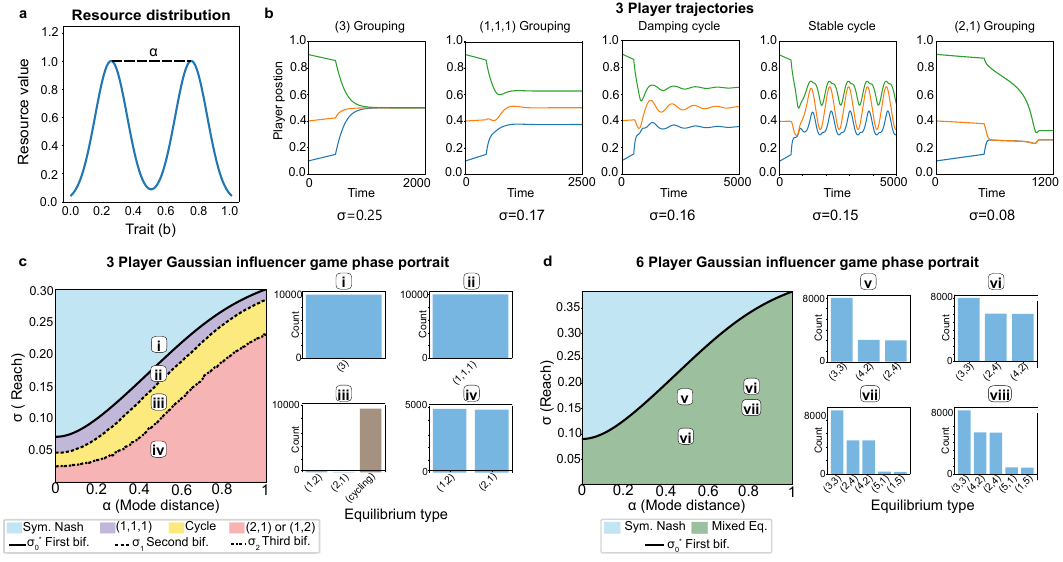}
\caption{\textbf{Rich dynamics in one-dimensional Gaussian influencer's games.} We study the canonical Gaussian kernel in one dimension, where stability reduces to a closed-form threshold for the first bifurcation, $\sigma_0^*$ (solid line in \textbf{c},\textbf{d}).
    \textbf{a},~Bimodal resource distribution $B(b)$ (equation~\eqref{eq:bimodal_resource}) with modes separated by $\alpha$ and bandwidth $\sigma_m$.
    \textbf{b},~Three-player trajectories at $\alpha = 0.5$ for $\sigma \in \{0.25,\, 0.17,\, 0.16,\, 0.15,\, 0.08\}$ (left to right). As the competitive reach $\sigma$ decreases, the stable symmetric Nash gives way to a dispersed $(1,1,1)$ equilibrium, then a limit cycle, and finally an asymmetric $(2,1)$ equilibrium.
    \textbf{c},~Bifurcation diagram for $N = 3$ across $\sigma \in (0, 0.3]$ (vertical) and $\alpha \in [0,1]$ (horizontal). Colors indicate equilibrium type: symmetric Nash (blue), dispersed $(1,1,1)$ (purple), limit cycles (yellow), asymmetric $(2,1)$ (pink). Curves mark the bifurcation boundaries $\sigma_0^*$ (solid), $\sigma_1$ (dotted) and $\sigma_2$ (dashed). Insets \textbf{i}--\textbf{iv} show equilibrium-type histograms at indicated $(\alpha,\sigma)$ values.
    \textbf{d},~Bifurcation diagram for $N = 6$, revealing additional equilibrium types including $(3,3)$, $(5,1)$ and $(4,2)$ clusters. Histograms \textbf{v}--\textbf{viii} confirm prevalence of each type across sampled initial conditions.
}\label{fig:1d_bifurcation}
\end{figure*}
In one dimension, equation~\eqref{eq:1d_stability} gives explicit conditions for stability and bifurcations. The Gaussian kernel,
\begin{equation}\label{eq:gaussian_kernel}
    f_i(x_i, b) = \exp\left(-\frac{(x_i - b)^2}{2\sigma^2}\right),
\end{equation}
where the standard deviation $\sigma$ is the competitive reach of influencer $i$, with smaller $\sigma$ creating more localized influence, is a natural choice given it is the one-dimensional equivalent of equation~\eqref{eq:MV_gaussian_kernel}. Univariate and multivariate Gaussian kernels capture smooth decay of influence with distance and are widely used across machine learning, ecology, and economics \cite{seeger2004gaussian,borger2008there,yang2020us}.

Moreover, the Gaussian kernel is log-concave and thus satisfies the existence and uniqueness conditions for a symmetric Nash equilibrium, which is given by the resource-weighted mean $x^* = \mathbb{E}_B[b] = \sum_{b \in \mathbb{B}} b\, B(b) \big/ \sum_{b \in \mathbb{B}} B(b)$. The stability condition (equation~\ref{eq:1d_stability}) reduces to a closed-form threshold $\sigma_0^*$ on the competitive reach $\sigma$:
\begin{equation}\label{eq:gaussian_stability}
    \sigma > \sigma_0^* := \sqrt{\frac{N-2}{N-1}}\,\sigma_B,
\end{equation}
where $\sigma_B$ is the standard deviation of the resource distribution. When the competitive reach $\sigma$ is above the threshold $\sigma_0^*$, the symmetric Nash equilibrium is locally stable. Below this threshold, the symmetric Nash equilibrium becomes unstable and a bifurcation cascade emerges. In Figure~\ref{fig:1d_bifurcation} we explore this structure for a bimodal resource distribution with varying mode separation $\alpha$ and fixed bandwidth $\sigma_m = 0.1$.

In Figure~\ref{fig:1d_bifurcation}\textbf{b} we show the sequence of bifurcations for three players at $\alpha = 0.5$ and initial positions $(0.1,0.4,0.9)$ as $\sigma$ decreases. Above the stability threshold $\sigma_0^*$, all trajectories converge to the symmetric Nash equilibrium $x^*=0.5$. Below $\sigma_0^*$, the symmetric Nash loses stability and a dispersed $(1,1,1)$ equilibrium emerges, in which all players occupy distinct positions between the two modes. This first bifurcation is a pitchfork (SI Section~3.2): in one dimension the two reduced eigenproblems of Theorem~\ref{thm:spectral} are scalar, so all Jacobian eigenvalues are real, and the symmetric equilibrium loses stability when a real eigenvalue crosses zero, with the $(1,1,1)$ branch emerging continuously from $x^*$.
As $\sigma$ decreases past a second boundary $\sigma_1 = 0.158$, the $(1,1,1)$ equilibrium loses stability through a Hopf bifurcation: a complex-conjugate pair of Jacobian eigenvalues crosses the imaginary axis (at $\sigma_1$ the spectrum is $\pm 62.3i$ and $-111.8$), rather than two branches of equilibria colliding. A limit cycle therefore appears below $\sigma_1$; the damped oscillations visible at $\sigma = 0.16$ in Figure~\ref{fig:1d_bifurcation}\textbf{b} are the approach to the still-stable $(1,1,1)$ equilibrium just above $\sigma_1$.
While cycling, players cyclically chase each other around the resource landscape. Finally, below a third boundary $\sigma_2$, a bistable asymmetric system occurs with $(2,1)$ and $(1,2)$ equilibria, in which two players cluster together around one mode and the third player occupies a distinct position on their own. Across $10{,}000$ random initial conditions at these parameters (Figure~\ref{fig:1d_bifurcation}\textbf{c}, insets \textbf{i}--\textbf{iv}), the outcomes are essentially unanimous: every run reaches the symmetric equilibrium at $\sigma = 0.25$ and the dispersed $(1,1,1)$ equilibrium at $\sigma = 0.18$, $97\%$ of runs cycle at $\sigma = 0.125$, and every run reaches a $(2,1)$ or $(1,2)$ equilibrium at $\sigma = 0.05$.

To study this behavior across resource parameters, we construct the bifurcation diagram in Figure~\ref{fig:1d_bifurcation}\textbf{c} for $N = 3$ across $\sigma \in (0, 0.3]$ and $\alpha \in [0,1]$. At $\alpha=0$ there is one mode, and at $\alpha=1$ the modes occur on the boundaries $0,1$. We see that the same bifurcation sequence occurs across the whole range of $\alpha$ values, so the bifurcation structure is not an artifact of a particular resource distribution. Moreover, the bifurcation boundaries $\sigma_0^*$, $\sigma_1$, $\sigma_2$ are all increasing in $\alpha$, indicating that as the modes become more separated, players need larger competitive reach $\sigma$ to maintain stability.
\begin{figure}[t]
\centering
\includegraphics[]{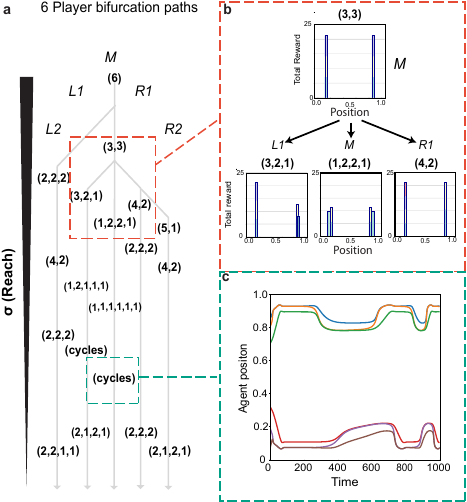}
\caption{\textbf{Sample of bifurcation behavior for six players.}
    \textbf{a},~Bifurcation paths for five initial conditions at $\alpha = 0.8$ as the competitive reach $\sigma$ decreases from $0.4$ to $0.03$.
    Identical parameters yield qualitatively different outcomes depending on initialization. Clusters of 5, 4, and 3 co-located players emerge and in turn bifurcate into smaller local games and limit cycles.
    \textbf{b},~Hierarchical bifurcation tree for three primary branches ($L1$, $M$, $R1$). Stacked bars show resource shares by cluster; local games reproduce the bifurcations of fewer-player games, illustrating how the global game decomposes as influence becomes more localized.
    \textbf{c},~Agent trajectories in the limit-cycle regime ($\sigma = 0.15$), showing cyclic behavior from three-player local games within the six-player game.
}\label{fig:bifurcation_paths}
\end{figure}

We construct the bifurcation diagram for $N = 6$ in Figure~\ref{fig:1d_bifurcation}\textbf{d}, which reveals additional equilibrium types including $(3,3)$, $(5,1)$, and $(4,2)$ clusters. The bifurcation structure is more complex than the three-player case, with multiple branches of equilibria emerging at different $\sigma$ values and coexisting over wide parameter ranges. Histograms \textbf{v}--\textbf{viii} give the prevalence of each equilibrium type across $20{,}000$ random initial conditions. At $\alpha = 0.5$ the $(3,3)$ equilibrium is reached from $60\%$ of initial conditions at $\sigma = 0.18$ and from $40\%$ at $\sigma = 0.10$, with the $(4,2)$ and $(2,4)$ equilibria taking the remainder; at $\alpha = 0.8$ the $(5,1)$ and $(1,5)$ equilibria appear as well, from $3\%$ of initial conditions at $\sigma = 0.20$ and $8\%$ at $\sigma = 0.15$. The same parameters thus host different equilibria, and the type reached depends on initialization.

While exploring the dynamics of five sampled initial conditions for six players (Figure~\ref{fig:bifurcation_paths}\textbf{a}), we see that clusters of players decompose into smaller local games as $\sigma$ decreases (we use ``local game'' to mean a spatially localized cluster of players that behaves as a smaller influencer's game, distinct from the standard game-theoretic notion of a subgame). These five branches are a sample, not an exhaustive tree: there are many other branches that we have not sampled, with several having flipped ordering, such as a $(2,4)$ equilibrium instead of a $(4,2)$ equilibrium; however, the local-game structures remain consistent across sampling, as we saw in Figure~\ref{fig:1d_bifurcation}\textbf{d} where we sampled the equilibrium types. Every sampled branch leaves the symmetric $(6)$ configuration at the same point, $\sigma = 0.341$, matching the closed-form threshold $\sigma_0^* = \sqrt{\tfrac{N-2}{N-1}}\,\sigma_B = 0.341$ for this resource distribution; the branches differ only in what happens below it. For example, in branch $M$ we see that a six-player game first decomposes into a $(3,3)$ equilibrium, which then bifurcates at $\sigma = 0.054$ into a $(1,2,2,1)$ equilibrium containing pairs of clusters $(1,2)$ and $(2,1)$ that arise from the two three-player local games. Under a different initial condition in R1 the six-player game bifurcates into a $(3,3)$ equilibrium first, then reorganizes into a $(4,2)$ equilibrium at $\sigma = 0.268$, before bifurcating into a $(2,2,2)$ dispersed equilibrium at $\sigma = 0.063$.
Figure~\ref{fig:bifurcation_paths}\textbf{b} shows the local-game decomposition as a hierarchical bifurcation tree for three primary branches ($L1$, $M$, $R1$). In $L1$ the two local games do not need to bifurcate at the same $\sigma$ value: one splits at $\sigma = 0.054$ while the other persists until $\sigma = 0.052$, explained by a difference in initial positions of the players. The stacked resource shares in Figure~\ref{fig:bifurcation_paths}\textbf{b} show that these configurations are far from payoff-equalizing. In the $(4,2)$ equilibrium of branch $R1$, for instance, the two clusters each take $50\%$ of the resource, since each sits on its own mode, which leaves $12.5\%$ for each of the four players at one mode and $25\%$ for each of the two players at the other. A player's rough payoff becomes one over the size of its cluster.

Finally, Figure~\ref{fig:bifurcation_paths}\textbf{c} shows the cyclic behavior of three-player local games within the six-player game in the limit-cycle regime ($\sigma = 0.15$). The complex dynamics of smaller-player games can therefore persist and interact within larger games as influence becomes more localized.

\begin{figure*}[t]
\centering
\includegraphics[width=\textwidth]{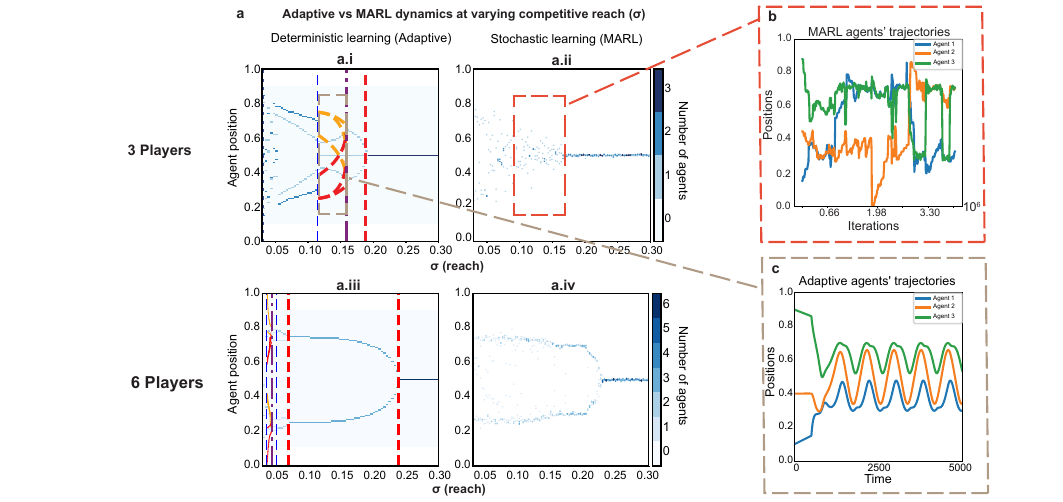}
\caption{\textbf{Bifurcation structure is robust under multi-agent reinforcement learning.}
    \textbf{a},~Bifurcation diagrams comparing gradient-ascent dynamics (left) and independent Q-learning (right) for 3 and 6 players show stability near $\sigma_0^*$ and give way to clustered equilibria. The resulting dynamics mimic the bifurcation structure of the adaptive dynamics, demonstrating a pitchfork bifurcation at $\sigma_0^*$. For the three-player dynamics, we observe a diffuse distribution of final player strategic positions, explained by cyclic behavior and the coexistence of $(2,1)$ and $(1,2)$ equilibria.
    \textbf{b},~Independent Q-learning agent trajectories for $N = 3$ at competitive reach $\sigma = 0.15$, exhibiting persistent collective oscillations that explain in part the diffuse distribution of final strategic positions in the bifurcation diagram.
    \textbf{c},~Corresponding adaptive-dynamics trajectories at the same parameters, which show the same cyclic attractor.}\label{fig:marl}
\end{figure*}

\paragraph{Reinforcement Learning Dynamics}
The analytical results above assume perfect payoff knowledge and continuous adjustment. Extending to realistic learning conditions, we study the same one-dimensional Gaussian influencer's games under independent Q-learning (IQL) \cite{watkins1992q,tan1993multi}, a decentralized multi-agent reinforcement learning (MARL) algorithm in which each agent maintains a private value table indexed only by its own strategic position (see Methods for full details). Compared with adaptive dynamics, three things are different. First, an agent sees only its own position and not its competitors', so many distinct configurations of the game look the same to it, a form of state aliasing \cite{mccallum1995instance}. Second, actions are chosen stochastically, through an $\varepsilon$-Boltzmann policy (Methods). Third, positions live on a discrete grid, which introduces a quantization error that continuous gradient ascent does not have.

Despite these differences, the bifurcation structure is qualitatively preserved.
In Figure~\ref{fig:marl}\textbf{a}, bifurcation diagrams comparing adaptive dynamics and IQL for 3 (Figures~\ref{fig:marl}\textbf{a.i,a.ii}) and 6 agents (Figures~\ref{fig:marl}\textbf{a.iii,a.iv}) show that the symmetric Nash equilibrium loses stability near the theoretical threshold~$\sigma_0^*$ in both cases, although the empirical bifurcation points differ systematically between the two dynamics (see below), and clustered equilibria emerge at lower competitive reach~$\sigma$. In the adaptive dynamics figures (Figures~\ref{fig:marl}\textbf{a.i,a.iii}) the agents' final positions are shown across varying $\sigma$ values ([0.03,0.3]); we show the envelopes of the cycle when present, indicated by the dashed yellow and red curves. Similarly, in the MARL figures (Figures~\ref{fig:marl}\textbf{a.ii,a.iv}) we utilize IQL to determine the final positions of agents at different values of $\sigma$; we box the approximate region of cycles in Figure~\ref{fig:marl}\textbf{a.ii}. Under adaptive dynamics the empirical thresholds coincide with the closed-form prediction to the resolution of the sweep ($\sigma = 0.188$ against $\sigma_0^* = 0.189$ for $N = 3$, and $0.238$ against $0.239$ for $N = 6$). Two systematic differences are visible between adaptive and IQL dynamics: IQL agents bifurcate at slightly lower competitive reach~$\sigma$ ($\sigma \approx 0.169$ against $0.188$ for $N = 3$, and $\approx 0.226$ against $0.238$ for $N = 6$, a shift of $\Delta\sigma \approx 0.01$--$0.02$; likely because the discrete grid stabilizes the symmetric equilibrium by limiting the precision with which agents can detect off-center incentives; SI Section~7.3), and IQL diagrams are more diffuse owing to residual exploration noise.

In the three-player case, Figure~\ref{fig:marl}\textbf{b} shows that the diffuse distribution of the final strategic positions after $4{,}000$ steps averaged over 75 trials is explained in part by persistent collective oscillations, which are also visible in the corresponding adaptive-dynamics trajectories (Figure~\ref{fig:marl}\textbf{c}). These cycles persist under both learning dynamics, so they are not an artifact of a particular learning algorithm.

\paragraph{Analytical extensions to higher dimensions and other kernels}
Because equation~\eqref{eq:general_depth} holds in any dimension, the stability analysis carries over directly to higher-dimensional strategy spaces, where $C$ and $S$ become $L\times L$ matrices rather than scalars. The multivariate Gaussian influence kernel (equation~\ref{eq:MV_gaussian_kernel}), used in the job market game, generalizes influence to the case where the covariance matrix $\Sigma$ controls competitive reach along each dimension. As in the 1D case where the symmetric Nash is the resource-weighted mean, in the multivariate case the symmetric Nash equilibrium is the resource-weighted centroid $x^* = \mathbb{E}_B[b]$. The 1D stability threshold generalizes: instability results from the direction of greatest resource spread. Specifically, the symmetric Nash equilibrium is locally stable if and only if
\begin{equation}\label{eq:mv_gaussian_stability}
    \sigma^2 > \frac{N-2}{N-1}\,\Lambda_{\max}(\Sigma_B) = (\sigma_0^*)^2,
\end{equation}
where $\Lambda_{\max}(\Sigma_B)$ is the largest eigenvalue of the resource covariance matrix; in one dimension $\Lambda_{\max}(\Sigma_B) = \sigma_B^2$, so this reduces exactly to the threshold in equation~\eqref{eq:gaussian_stability}. This prediction is consistent with the empirical job market of Figure~\ref{fig:linkedin}: the job-posting distribution has $\Lambda_{\max}(\Sigma_B) = 0.18$, so equation~\eqref{eq:mv_gaussian_stability} predicts a threshold $\sigma_0^* = 0.43$ at $N = 1{,}000$, bracketed by the transition between Figure~\ref{fig:linkedin}\textbf{c.i} and \textbf{c.ii}, where job seekers occupy a single cluster at $\sigma = 0.45$ and have split into multiple clusters by $\sigma = 0.24$.
In the job market game, this implies that the job opportunity spread is what dictates why job seekers cluster around certain skill profiles, and as the spread of job opportunities increases, a job seeker will need to be increasingly generalist to be competitive across job traits. Notably, if a job market is not varied enough relative to the job seeker's breadth of abilities  (their $\sigma>\sigma^*_0$) then specialist behavior does not emerge and job seekers cluster around the centroid of the job market, leaving opportunities away from the centroid poorly served.

\begin{figure}[t]
\centering
\includegraphics[]{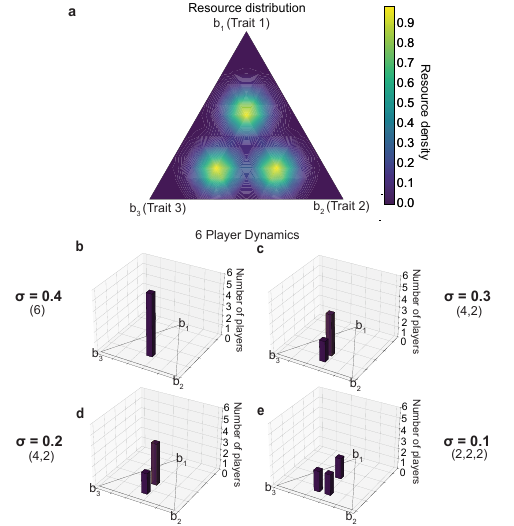}
\caption{\textbf{Dirichlet influencer's games on the simplex: niche partitioning in higher-dimensional trait spaces.}
    \textbf{a},~Resource distribution over a $2$-simplex resource trait space; color encodes resource value (yellow, high; purple, low).
    \textbf{b},~Equilibrium strategic positions for $N = 6$ players at competitive reach $\sigma=0.4$. Players form a single cluster of six, the $(6)$ configuration, at the symmetric Nash equilibrium.
    \textbf{c},~Equilibrium strategic positions at competitive reach $\sigma=0.3$; players bifurcate into a $(4,2)$ configuration.
    \textbf{d},~Equilibrium strategic positions at competitive reach $\sigma=0.2$; players remain in the $(4,2)$ configuration.
    \textbf{e},~Equilibrium strategic positions at competitive reach $\sigma=0.1$; players bifurcate into a $(2,2,2)$ configuration.
    As competitive reach narrows, the same bifurcation sequence identified in one dimension emerges despite the change in resource space and influence kernel family. Larger clusters give way to progressive clustering around resource modes, with local games reproducing the equilibrium types of lower-player games.}\label{fig:2d_dynamics}
\end{figure}

We also analyze the Dirichlet and Beta influence kernels, defined on a simplex strategy space, parameterized so that the strategic position $x_i$ is the mode with a reach parameter $0<\sigma$. The Dirichlet kernel is
\begin{equation}
            \label{eq:dirichlet_mode_kernel}
            f_i(x_i, b;\sigma) = \frac{\Gamma(L+\frac{1}{\sigma})}{\prod_{l=1}^L \Gamma(1+\frac{1}{\sigma} x_{(i,l)})}
            \prod_{l=1}^L b_l^{\,\frac{1}{\sigma} x_{(i,l)}}, \qquad b,x_i \in \Delta^{L-1}
\end{equation}
where $L$ is the number of dimensions, $\Gamma$ is the gamma function, and $x_{(i,l)}$ is the $l$-th component of player $i$'s strategic position. One key difference is that the Dirichlet kernel is defined on a bounded domain and adaptive dynamics rely on gradient projection. However, as demonstrated in Figure~\ref{fig:2d_dynamics}, the same qualitative bifurcation phenomena emerge as the competitive reach decreases: players cluster around resource peaks and local games reproduce the equilibrium types of smaller contests. Unlike the Gaussian and multivariate Gaussian cases, the location of the symmetric Nash equilibrium depends on the competitive reach $\sigma$ for the Dirichlet and Beta kernels. Details are provided in the SI (Section~4.2).

These extensions all work because the game is symmetric: the Jacobian at the symmetric equilibrium is block-circulant, so its $NL \times NL$ spectrum splits into two $L \times L$ eigenproblems that do not grow with $N$ (Methods; SI). This reduction holds for any symmetric influencer's game with a quasi-concave influence kernel (Theorem~\ref{thm:block_jacobian}), not only the Gaussian studied in detail here. Both the Beta (SI Section~3.3) and Dirichlet (SI Section~4.2) kernels exhibit qualitatively identical bifurcation structure, and independent Q-learning agents reproduce the same phenomena.

\section{Discussion}
Every symmetric influencer's game with a strictly log-concave influence kernel possesses a unique symmetric Nash equilibrium, and this equilibrium is stable under gradient-ascent dynamics whenever there are two players. When the symmetric equilibrium does destabilize, the game transitions to a nested hierarchy of local influencer's games, each reproducing the equilibrium structure of a smaller contest. Niche structure thus emerges from competitive reach, the number of players, and the shape of the resource distribution alone, without further assumptions on player heterogeneity or external forcing.

Set against Hotelling's spatial competition model and other preexisting spatial competition models, the value of the influencer's game is that our model can describe much of the preexisting models' behaviors. For instance, Hotelling's duopoly and the facility-location literature established the pull toward minimum differentiation; that behavior is the broad-reach regime for an influencer's game for any number of players. Importantly, for two players it holds at every reach, broadening the 2-player minimum differentiation result to a stability one, since two players with strictly log-concave kernels converge to the symmetric equilibrium in any dimension (Theorem~\ref{thm:2playerstability}). Whereas in spatial voting models results like the mean and median voter theorems, reappear in the influencer's game as limiting cases rather than as individual constructions (SI Section~5.2.2). While, Colonel Blotto games treat allocation as an all-or-nothing contest between two players over discrete battlefields; the influencer's game's proportional allocation rule relaxes all three of those restrictions at once. Thus allowing the influencer's game to admit arbitrarily many players, continuously varying allocation, and a resource density that varies over a space of any dimension (SI Section~6). Additionally, the influencer's game's allocation rule mirrors the Tullock contest success function \cite{tullock1980efficient,skaperdas1996contest}, but applied at every point of a continuum of resource points and a position-dependent influence in place of effort. Finally from ecological niche theory, resource partitioning sustains coexistence; in the influencer's game partitioning emerges endogenously, as in evolutionary-branching models of niche formation \cite{dieckmann1999origin}, from competitive reach, player number, and resource shape.

For strictly log-concave kernels, existence and uniqueness hold at every competitive reach and every player count. Whether competition produces specialization therefore depends only on whether the symmetric equilibrium is stable, and the stability condition has a closed form: $\sigma_0^* = \sqrt{(N-2)/(N-1)}\,\sigma_B$ in one dimension and equation~\eqref{eq:mv_gaussian_stability} in general. Independent Q-learning agents reproduce the same equilibrium types and bifurcate within $\Delta\sigma \approx 0.02$ of the analytical threshold (SI Section~7.3); and the threshold is consistent with an empirical landscape: it predicts $\sigma_0^* = 0.43$ for the LinkedIn job-posting distribution, which lies between the single cluster observed at $\sigma = 0.45$ and the multiple clusters observed at $\sigma = 0.24$.

When applied to labor markets, the influencer's game indicates, whether through more precise skill screening or narrower job requirements, when effective job seeker competitive reach decreases the result is workers being pushed toward increasingly specialized occupational profiles. This occupational specialization is a pattern that is at least qualitatively consistent with evidence on skill-biased technical change and occupational polarization \cite{autor2003skill,acemoglu2011skills,goos2007lousy,autor2013growth}. In electoral competition, the same threshold determines whether candidates converge to the electorate's center or fragment into polarized niches. This extends the endogenous polarization mechanisms of Jones et al.\ \cite{jones2022polarization} (asymmetric voter distributions, fixed third parties, abstaining voters) to any number of active candidates, without exogenous preference shocks. Within the SI (Section~5.2) we develop the connection to political science in more detail. By adding an abstaining vote option and a fixed third player in the influencer game we are able to reproduce the destabilizing effects of fixed third parties and abstaining voters within our framework. In digital ecosystems such as social media, the local-game structure offers a game-theoretic explanation of content niches and filter bubbles, especially as algorithms effectively narrow each creator's competitive reach \cite{flaxman2016filter,bakshy2015exposure,cinelli2021echo,bak2021stewardship}.

Several areas around influencer games still need to be developed. The stability results are limited to local behavior. We have not precisely described the geometry of basins of attraction for the coexisting asymmetric equilibria and limit cycles (Figures~\ref{fig:1d_bifurcation} and~\ref{fig:bifurcation_paths}), which remain analytically open. The largest assumption that we make is that the game is symmetric, which in turn limits some of the applicability of the results. A next step would be extending the theory to heterogeneous players, or to hierarchical structures with players operating at different spatial scales. Two additional open problems include varying resource distributions (capturing seasonal or trend-driven shifts) and coalition formation.

\section{Methods} \label{sec:methods}

Stability of the symmetric equilibrium $\mathbf{x}^*$ is studied under gradient-ascent dynamics $\nabla_t x_i = \nabla_{x_i} u_i(\mathbf{x})$. By Lyapunov's linearization theorem \cite{hirsch2013differential}, local stability is determined by the eigenvalues of the Jacobian matrix $\mathbf{J}_{\mathbf{u}}(\mathbf{x}^*) \in \mathbb{R}^{NL\times NL}$. The symmetry of the game endows this Jacobian with a block-circulant structure (Theorem~\ref{thm:block_jacobian}), which admits a Kronecker-product decomposition \cite{horn2012matrix} that reduces the $NL\times NL$ spectral problem to two $L\times L$ eigenproblems:
\begin{equation}
    spec(\mathbf{J}_{\mathbf{u}}) = spec(A^{+}) \cup spec(A^{-}),
\end{equation}
where $A^{+} := A + (N-1)\Delta$ and $A^{-} := A - \Delta$, with $A$ and $\Delta$ denoting the self- and cross-interaction Hessian blocks (Theorem~\ref{thm:spectral}; proof in SI Section~2.3). The symmetric equilibrium is locally asymptotically stable under gradient-ascent dynamics if and only if both $A^+$ and $A^-$ are negative definite (away from the threshold itself).

Both $A^+,A^-$ separate the effect of a player's own payoff curvature ($A$) from the influence of competitors ($\Delta$). This makes the cause of instability clearer, and we notice that a bifurcation of a symmetric equilibrium occurs when the pull of competitors outweighs the curvature of a player's own payoff landscape. Expressed through the curvature $C$ and scatter $S$ of the log-kernel, the blocks are $A^{+} = -\frac{N-1}{N^2}C$ and $A^{-} = -\frac{N-1}{N^2}\bigl(C - \frac{N-2}{N-1}S\bigr)$, which yields the stability condition of equation~\eqref{eq:general_depth} in any dimension (Theorem~\ref{thm:general_stability}; proof in SI Section~2.3.3). For two players the scatter term vanishes. Due to strictly log-concave kernels making $C$ positive, the symmetric Nash equilibrium is locally stable in every dimension (Theorem~\ref{thm:2playerstability}; proof in SI Section~2.3). For $L = 1$ and $N > 2$, the stability condition becomes (Corollary~\ref{corr:1d_stability})
\begin{equation}
    \sum_{b \in \mathbb{B}} B(b) d'(b)<-\frac{N-2}{N-1}\sum_{b \in \mathbb{B}} B(b) d(b)^2,
\end{equation}
where $d(b)=\partial_{x_i} \ln f_i(x^*, b)$ and $d'(b) = \partial_{x_i}^2 \ln f_i(x^*, b)$.

\paragraph{Adaptive dynamics}
In our adaptive dynamics model, all players simultaneously adjust their strategic positions via gradient ascent on their payoff functions. At each time step every agent takes a step in the direction of its payoff gradient:
\begin{equation}
    x_i(t+1) =x_i(t) + \eta \cdot \nabla_{x_i} u_i(\mathbf{x}).
\end{equation}
where $\eta$ is the learning rate. The gradient of the payoff function $u_i(\mathbf{x})$ with respect to $x_i$ is denoted by $\nabla_{x_i} u_i(\mathbf{x})$ which is $L$-dimensional. This idealized learning mechanism assumes that agents have perfect knowledge of their payoff gradients and can adjust their strategic positions continuously. Using static learning rates (listed in the SI) for given competitive reach $\sigma$, agents converge to locally stable equilibria, such as the symmetric Nash equilibrium, or to limit cycles and asymmetric equilibria when the symmetric equilibrium is unstable.

\paragraph{Reinforcement learning dynamics.}
We also employ independent Q-learning dynamics to test the bifurcation structure under partial observability, stochastic action selection and discretized strategy spaces. Each agent maintains a Q-table representing the value it learns for each state--action pair from interactions with the environment (the influencer's game), updated via
\begin{equation}
    Q_i(s, a) \leftarrow Q_i(s, a) + \eta_{rl}\bigl[r + \gamma \max_{a'} Q_i(s', a') - Q_i(s, a)\bigr],
\end{equation}
where $s$ is the current state, $a$ is the action taken, $r$ is the reward, $\eta_{rl}$ is the learning rate and $\gamma$ is the discount factor. Here we denote the new state and actions as $s',a'$ after the current action was made in the current state. Given that agents learn independent value functions $Q$, without knowledge of their opponents, the independent Q-learning approach represents a partially-observable learning environment.

Action selection follows an $\varepsilon$-Boltzmann policy: with probability $\varepsilon$ the agent explores by choosing a random action, and with probability $1-\varepsilon$ it exploits by sampling from a Boltzmann (softmax) distribution over its Q-values:
\begin{equation}
     P(a|s) = \frac{e^{Q(s,a)/\tau}}{\sum_{a'} e^{Q(s,a')/\tau}}
\end{equation}
where $\tau$ is a temperature parameter. The temperature $\tau$ controls how much the agent explores, with choices close to uniform at large $\tau$ and close to greedy at small $\tau$. This softmax rule improves learning stability by preventing premature convergence, which is a particular risk when the payoff landscape has several local optima \cite{sutton2018reinforcement}. Full learning parameters are reported in the SI.

\paragraph{Empirical analysis of data on 2d job market sentiment.}
To ground the framework empirically, we analyze a dataset of LinkedIn job postings from 2023--2024 \cite{arsh_koneru_2024}. Using the job posting description, each posting is scored by the language model GPT-OSS-20B (temperature $0.1$) on 0--1 technology and finance scales anchored by category descriptions (prompt in SI Section~5.1.1). The resulting two-dimensional score from GPT-OSS-20B is the job posting's trait position $(b_1,b_2)$ where $b_1$ and $b_2$ are the technology and finance scores, respectively. We retain postings with a score of at least $0.1$ in one of the two categories, bin them on a $5 \times 5$ grid over $\mathbb{B} = [0,1]^2$, and take the count of postings in each cell as the resource $B(b)$ at the cell center, with no smoothing. We then simulate adaptive dynamics for symmetric job seekers with multivariate Gaussian influence kernels ($\Sigma=\sigma^2 I_{2}$) under varying competitive reach values $\sigma$.

\paragraph{Bifurcation analysis (adaptive dynamics).}
To map the full bifurcation structure we vary both the competitive reach $\sigma$ and the mode separation $\alpha$ of the one-dimensional resource distribution, taken to be a symmetric bimodal Gaussian mixture:
\begin{equation}\label{eq:bimodal_resource}
    B(b) = e^{-(b - \mu_1)^2/2\sigma_m^2} + e^{-(b - \mu_2)^2/2\sigma_m^2}
\end{equation}
where $\mu_1 = 0.5 - \alpha/2$, $\mu_2 = 0.5 + \alpha/2$, and $\sigma_m$ controls the width of each mode. For Figures showing trajectories at fixed $\alpha$ (Figures~\ref{fig:linkedin}\textbf{b}, \ref{fig:bifurcation_paths}, and~\ref{fig:2d_dynamics}), $\sigma$ is swept with initial positions held fixed. The boundaries in the $(\sigma, \alpha)$ diagrams (Figure~\ref{fig:1d_bifurcation}\textbf{c},\textbf{d}) are obtained in the same way: at each of $100$ ($N = 3$) or $200$ ($N = 6$) values of $\alpha \in [0,1]$, $\sigma$ is swept from a fixed initial condition and the values at which the outcome type changes are recorded, while $\sigma_0^*$ is drawn from its closed form. At the $(\sigma, \alpha)$ points shown as insets we additionally sample $10{,}000$ ($N = 3$) or $20{,}000$ ($N = 6$) initial conditions uniformly on $[0,1]^N$ and classify each run according to the outcome of gradient-ascent dynamics.

Each run is first tested for convergence: a trajectory is deemed to have reached a fixed-point equilibrium if every agent's position changes by less than $10^{-8}$ ($N = 3$) or $10^{-12}$ ($N = 6$) over two consecutive steps within $3{,}000$ ($N = 3$) or $6{,}000$ ($N = 6$) steps. A run that has not converged is then tested for a cycle. We drop the first $1{,}000$ steps, keep the last $500$ positions of one agent, and cut them into consecutive non-overlapping windows of length $p$, trying each odd $p$ from $5$ to $49$. We call the run a limit cycle if the absolute Pearson correlation between the first two windows is above $0.5$ and its average with the third window, when there is one, is above $0.6$. There are also trajectories that neither converged to a fixed point nor represented cycles. We exclude these from the count. Of the trajectories that could not be classified, none occurred in the $N = 3$ samples and one in $20{,}000$ in the $N = 6$ samples reported in Figure~\ref{fig:1d_bifurcation}. If a run did converge those runs were further classified by grouping final agent positions that lie within $10^{-3}$ of their nearest neighbor. Those close enough to one another were classified into clusters and then named based on the resulting cluster sizes (symmetric Nash, dispersed, or asymmetric cluster).

\paragraph{Bifurcation analysis (independent Q-learning).}
To test whether the bifurcation structure survives independent Q-learning, we replicate the one-dimensional Gaussian sweeps using $N=3,6$ independent Q-learning agents in place of gradient ascent. To start the strategy space which is originally a continuous 1D line $[0,1]$ is replaced by a grid of 11 points spaced $0.1$ apart. There are then three available actions to agents, first move left, second stay, or third move right. Thus when counting the number of states in each agent's Q-table the total is $11 \times 3$ (indexed by the position and action). We use a $\varepsilon$-Boltzmann policy that selects a uniformly random action with probability $\varepsilon = 0.1$ or otherwise samples from the softmax of the Q-values at temperature $\tau = 0.5$. The agent's Q-values are then updated by one-step Q-learning with learning rate $\eta_{rl} = 0.05$ and with a reward discount factor $\gamma = 0.5$. Finally we set the resource landscape to a bimodal Gaussian mixture with modes at $\mu_1=0.25$ and $\mu_2=0.75$ ($\sigma_m=0.1$), composed of 100 evenly spaced trait values.

During the experiment procedure we sweep the competitive reach over 96 values $\sigma\in[0.03,\,0.3]$. For each of these 96 $\sigma$ values, we initiate 100 independent trials each with 4{,}000 steps. To further refine the Q-tables we set them to persist across successive trials so that agents can refine their policies incrementally. To keep stability of the Q-values, the first 25 trials are discarded as burn-in trials.  We then averaged per agent the final positions from the remaining 75 trials. The result is the IQL bifurcation diagram (Figure~\ref{fig:marl}).

\backmatter

\bmhead{Supplementary information}

Supplementary Information is available for this paper. It contains the full model formulation and measure-theoretic foundations (SI Section~1); complete proofs of existence, uniqueness, and stability theorems (SI Section~2); extended one-dimensional analysis including Beta influence kernels (SI Section~3); multi-dimensional strategy spaces with multivariate Gaussian and Dirichlet kernels (SI Section~4); applications to social media, to political science (voting models with fixed third parties and abstaining voters, and recovery of the median voter theorem), and to ecology (dispersal with movement costs) (SI Section~5); a formal construction relating influencer's games to Colonel Blotto games (SI Section~6); and full MARL training details including Q-learning specifications and convergence analysis (SI Section~7).

\bmhead{Acknowledgements}

We thank Alex McAvoy, Soroush Vousghi, Thomas Zdyrski, and Jonathan Lindbloom for their extensive feedback and for helpful discussions.

\section*{Declarations}

This research received no specific grant from any funding agency in the public, commercial, or not-for-profit sectors. The authors declare no competing interests. Ethics approval, consent to participate, and consent for publication are not applicable to this study, and no new materials were generated. The LinkedIn job-posting dataset is publicly available \cite{arsh_koneru_2024}, and the processed resource landscape and the simulation outputs underlying the figures are available in the InflGame repository (\url{https://github.com/MarksLovett/InflGame}), which also hosts the open-source InflGame Python package used for all simulations. M.L. and F.F. designed the study, M.L. developed the theory, software, and simulations, and M.L. and F.F. wrote the manuscript.

\noindent

\begin{appendices}
\section{Formal Theorem Statements}\label{app:theorems}

This appendix collects the formal statements of the main theoretical results. Full proofs are provided in the Supplementary Information: existence and uniqueness of Nash equilibria (SI Section~2.1--2.2), stability analysis, spectral decomposition, and the general stability condition (SI Section~2.3), Gaussian equilibrium and stability threshold (SI Section~3.1).

\begin{theorem}[Existence of Symmetric Nash Equilibrium]\label{thm:existence}
    Let $T$ be a symmetric influencer's game with $\mathbb{D}$ compact and convex and influence kernels $f(x_i, b)$ quasi-concave in~$x_i$. Then there exists at least one symmetric Nash equilibrium $\mathbf{x^*}=(z^*, \dots, z^*)\in \mathbb{D}$.
\end{theorem}

\begin{theorem}[Uniqueness of Symmetric Nash Equilibrium]\label{thm:uniqueness}
    Let $T$ be a symmetric influencer's game with $\mathbb{D}$ compact and convex and influence kernels $f(x_i, b)$ strictly log-concave in~$x_i$. Then the symmetric Nash equilibrium $\mathbf{x^*}$ is unique.
\end{theorem}

\begin{theorem}[Block Structure of the Jacobian]\label{thm:block_jacobian}
    Let $T$ be an $N$-player symmetric influencer's game with $\mathbb{D}$ compact and convex and influence kernels $f(x_i, b)$ quasi-concave in~$x_i$. Then the Jacobian of the adaptive dynamics at the symmetric Nash equilibrium $\mathbf{x^*}$ has the block structure
    \begin{equation}
        \mathbf{J}_\mathbf{u}(\mathbf{x}^*) = \begin{bmatrix}
        A & \Delta & \cdots & \Delta \\
        \Delta & A & \cdots & \Delta \\
        \vdots & \vdots & \ddots & \vdots \\
        \Delta & \Delta & \cdots & A
        \end{bmatrix},
    \end{equation}
    where $A\in \mathbb{R}^{L \times L}$ is the self-interaction Hessian block and $\Delta \in \mathbb{R}^{L \times L}$ is the cross-interaction Hessian block.
\end{theorem}

\begin{theorem}[Spectral Decomposition and Stability Reduction]\label{thm:spectral}
    Let $A^{+} := A + (N-1)\Delta$ and $A^{-} := A - \Delta$. Then $spec(\mathbf{J}_{\mathbf{u}}) = spec(A^{+}) \cup spec(A^{-})$, reducing the $NL \times NL$ stability problem to two $L \times L$ eigenproblems independent of~$N$. The symmetric Nash equilibrium is locally asymptotically stable under gradient-ascent dynamics if and only if both $A^+$ and $A^-$ are negative definite.
\end{theorem}

\begin{theorem}[Stability in Arbitrary Dimension]\label{thm:general_stability}
    Let $T$ be a symmetric $N$-player influencer's game with $\mathbb{D} \subset \mathbb{R}^L$, influence kernels $f(x_i,b)$ twice continuously differentiable in~$x_i$, and an interior symmetric Nash equilibrium $\mathbf{x}^*$. Then $A^{+} = -\frac{N-1}{N^2}\,C$ and $A^{-} = -\frac{N-1}{N^2}\bigl(C - \frac{N-2}{N-1}S\bigr)$, where $C = -\sum_{b \in \mathbb{B}} B(b)\,\nabla_{x}^{2}\ln f(x^*,b)$ and $S = \sum_{b \in \mathbb{B}} B(b)\,\nabla_{x}\ln f(x^*,b)\,\nabla_{x}\ln f(x^*,b)^{\top}$. If $C$ is positive definite, $\mathbf{x}^*$ is locally asymptotically stable under gradient-ascent dynamics when $\rho < 1$ and unstable when $\rho > 1$, with $\rho$ as in equation~\eqref{eq:general_depth}.
\end{theorem}

\begin{theorem}[Two-Player Stability]\label{thm:2playerstability}
    Let $T$ be a 2-player symmetric influencer's game with $\mathbb{D}$ compact and convex and influence kernels $f(x_i, b)$ strictly log-concave in~$x_i$. Then the symmetric Nash equilibrium $\mathbf{x^*}$ is locally asymptotically stable under gradient-ascent dynamics.
\end{theorem}
Theorem~\ref{thm:2playerstability} is the case $N = 2$ of Theorem~\ref{thm:general_stability}, for which $\rho = 0$.

\begin{corollary}[One-Dimensional Stability Condition]\label{corr:1d_stability}
    For a symmetric $N$-player influencer's game with $L = 1$, the symmetric Nash equilibrium is locally asymptotically stable if and only if
    \begin{equation}
        \sum_{b \in \mathbb{B}} B(b) \bigl[ (N-1)\, d'(b) + (N-2)\, d(b)^2 \bigr] < 0,
    \end{equation}
    where $d(b) = \partial_{x_i} \ln f_i(x^*, b)$ and $d'(b) = \partial_{x_i}^2 \ln f_i(x^*, b)$.
\end{corollary}

\begin{lemma}[Symmetric Nash Equilibrium for Gaussian influencer's games]\label{lemma:gaussian_equilibrium}
    For a symmetric $N$-player influencer's game with a Gaussian influence kernel, the symmetric Nash equilibrium is
    \begin{equation}
        x^* = \frac{\sum_{b \in \mathbb{B}} B(b)\, b}{\sum_{b \in \mathbb{B}} B(b)} = \mathbb{E}_{B}[b].
    \end{equation}
\end{lemma}

\begin{corollary}[Gaussian Stability Threshold]\label{cor:gaussian_stab}
    The symmetric Nash equilibrium $x^*$ of the Gaussian influencer's game is locally asymptotically stable if and only if
    \begin{equation}
        \sigma > \sigma_0^* := \sqrt{\frac{N-2}{N-1}}\;\sigma_B, \qquad \sigma_B = \sqrt{\mathrm{Var}_B(b)}.
    \end{equation}
\end{corollary}

\end{appendices}

\bibliography{sn-bibliography}%
\end{document}